\documentclass[12pt,a4paper]{article}
\usepackage[pdftex,unicode]{hyperref}

\usepackage{mathtext}
\usepackage[T2A]{fontenc}
\usepackage[utf8]{inputenc}
\usepackage[russian,english]{babel}

\usepackage{amsmath,amssymb}
\usepackage{amsxtra}
\usepackage{color}
\usepackage{array}
\usepackage{verbatim}
\usepackage{appendix}
\usepackage{authblk}
\usepackage{graphicx}

\newcommand{\ZZ}{\mathbb{Z}}
\newcommand{\RR}{\mathbb{R}}

\newcommand{\gscm}{\text{g}/\text{cm}^2}
\newcommand{\gccm}{\text{g}/\text{cm}^3}
\newcommand{\eps}{\varepsilon}
\newcommand{\prob}{{\bf P}}
\newcommand{\me}{{\bf E}}
\newcommand{\var}{{\bf Var}}
\newcommand{\heaviside}{\Theta}
\newcommand{\kap}{{\text{\it\ae}}}
\newcommand{\config}{\mathfrak{A}}
\newcommand{\statves}{\mathfrak{g}}
\newcommand{\wigner}[3]{\left(\begin{array}{ccc}#1&#2&#3\\0&0&0\end{array}\right)}
\newcommand{\ii}{\mathfrak{i}}
\newcommand{\xxi}{\boldsymbol{\xi}}
\renewcommand{\exp}[1]{{\rm e}^{#1}}

\begin{document}

\title{Fourier Transform Method of a Detailed Configuration Accounting in Hot Plasma Bound-Bound Opacity Calculations}

\author{E.~Yu.~Arapova$^1$, Yu.~V.~Koryakina$^{1,2}$, M.~A.~Vronskiy$^{1,2}$\footnote{Corresponding author. E-mail: mavronskiy@vniief.ru, mavronskii@mephi.ru}}
\date{
\parbox{13cm}{
{\small\it
$^1${Russian Federal Nuclear Center --- All-Russian Research Institute of\\Experimental Physics}\\
$^2${Sarov Physics and Technology Institute branch of the National Research\\Nuclear University MEPhI (Moscow Engineering Physics Institute).\\SarPhTI NRNU MEPhI}
}
}
}
\maketitle

\begin{abstract}

G.~Hazak and J.~Kurzweil discovered a method of configurational resolution of transition arrays for the Super Transition Arrays approach to the bound-bound opacity calculation. Their method is based on the representation of the photoabsorption coefficient as the Fourier transform, the linearity of the transition energy between configurations with respect to shell occupation numbers, and factorization of the probabilities of configurations on shell occupation numbers. We extend the Hazak --- Kurzweil method for the calculations with Detailed Configuration Accounting. The resulting expressions for the bound-bound opacity represent an alternative to the widely used ones and are quite convenient for numerical implementation.

\end{abstract}

\section{Introduction}
The radiative opacity of hot plasma is the quantity that significantly affects various processes at high energy densities. Theoretical models and numerical algorithms for calculating opacity are the key subject of a large number of works (see \cite{Nikiforov2005}, \cite{Bauche2015} and references therein). 
An important class of physical models for opacity calculation are the average-atom models that allow to obtain average-atom potential, atomic cell electron density, one-electron states as a zero-order approximation for the plasma electron characteristics. Average-atom features are then used to get transition probabilities, electron configuration energies, transition energies, line widths, etc. These values give the data enabling to compute the bound-bound opacity, that is essential part of overall opacity. Bound-bound opacity is formed by the transitions that correspond to the jump of an electron between a pair of shells. Such a pair of shells and an initial configuration specified by the shell occupation numbers define a transition array having a detailed structure. This detailed structure description is often performed effectively within the framework of Unresolved Transition Arrays (UTA, see \cite{Bauche2015}, \cite{BarShalom1995}). Namely, to describe the mean value and dispersion of the energies of the transition array, an additional spectral shift and Gaussian width are introduced. 

Different initial configurations can bring to transition arrays with substantially different parameters, primarily the energies. The diversity of configurations, i.~e.\ the shell occupation numbers fluctuations, must be taken into account to obtain adequate opacity results. At the same time, the total number of configurations can be tremendous (e.\ g.\ \cite{Krief2021}), what makes their straight account impossible. Common approaches to include the impact of the shell occupation numbers fluctuations on the opacity are:
\begin{itemize}
\item replacement of transitions from different configurations with transition from an average ``configuration'' with an additional effective width describing the variance of transition arrays from different configurations (see \cite{Nikiforov2005}, \cite{Dragalov1987});
\item STA approach: combining configurations into superconfigurations what allows their enumerating, and replacing transitions arrays with Super Transitions Arrays with additional effective width describing the variance of transition arrays coresponding to supertransition (e.\ g.\ \cite{Bauche2015}, \cite{BarShalom1989}, \cite{Ovechkin2014}, \cite{Pain2021});
\item direct accounting of the most probable configurations with omitting the unlikely ones (e.\ g.\ \cite{Gaffney2011}).
\end{itemize}

In the series of papers \cite{Hazak2012}, \cite{Kurzweil2013}, \cite{Kurzweil2016} G.~Hazak and Y.~Kurzweil developed the method to obtain the configurationally resolved STA spectra that refines the impact of the shell occupations numbers fluctuations on the super transition array characteristics. Their method is based on three assumptions. The first is the possibility to write the bound-bound opacity as a Fourier transform. The second is the linearity of transition frequency with respect to shell occupation numbers. The third is the binomial approximation of the configuration probability, what corresponds to the Gibbs distribution on the set of configurations with average-atom zeroth approximation for the configuration energy. Essential component of the method are the recurrence relations for the introduced ``Complex Pseudo-Partition Functions''. 

This method can be extended to the Detailed Configuration Accounting (DCA) approach. The resulting extension leads to simple closed expressions for the required values.
The effective UTA description of the transition arrays detailed structure can be included in the proposed method due to that UTA widths are expressed through diagonal quadratic forms w.r.t.\ shells occupation numbers (see \cite{BarShalom1995}, \cite{Kurzweil2013}). We also introduce a probabilistic representation of some necessary quantities, that clarifies the essence of the method: the mathematical expectation of the product of independent random variables is equal to the product of the expectations of these random variables. The considered variant of DCA reduces the summation of a large number of terms to the calculation of the Fourier transform of more easily calculated expressions. This explains the proposed name of the method.

\section{Assumptions}
Suppose we are given spherically symmetric potential $V(x)$ and the corresponding set of bound electron shells (energies and normalized two-component Dirac radial wavefunctions)
\begin{equation}
\eps_j,\ (f_j(x),g_j(x)),\ \int_0^{\infty}(f_j^2(x)+g_j^2(x))dx = 1,\ j=1,\ldots,M.
\end{equation}
Every index $j$ corresponds to Dirac quantum number $\kap_j\in\ZZ\setminus\{0\}$ and statistical weight $\statves_j = 2|\kap_j|$. Denote by 
\begin{equation}
\config=\prod_{j=1}^{M}\{0,1,\ldots,\statves_j\}=\Bigl\{
{\bf n}=(n_1,n_2,\ldots,n_M),\quad 0\leq n_j\leq\statves_j,\ j=1,\ldots,M\Bigr\}
\end{equation}
the set of all electronic configurations. Here $n_j$, $j=1,\ldots,M$ are the shells occupation numbers. For every configuration ${\bf n}\in\config$ define the energy of the zeroth-order approximation --- independent electrons in the effective central field
\begin{equation}\label{ZeroEnergy}
E^{(0)}({\bf n})=\sum_{j=1}^Mn_j\eps_j
\end{equation}
and average configuration energy including the Coulomb interaction of the bound electrons unaccounted in the zeroth-order approximation
\begin{equation}\label{FirstEnergy}
E({\bf n})=\sum_{j=1}^Mn_jq_j+\frac 12\sum_{1\leq j,k\leq M}n_j(n_k-\delta_{kj})\theta_{jk}
\end{equation}
where 
\begin{equation}
q_j=\eps_j+\int_0^{\infty}R_{jj}(x)\left(V(x)-\frac{Ze^2}{x}\right)dx
,\quad \theta_{jk}=\frac{\statves_j}{\statves_j-\delta_{jk}}\left(F_{jk}^{(0)}-\frac 14\sum_{s=0}^{\infty}
\mathfrak{R}_{s\kap_j\kap_k}G_{jk}^{(s)}\right)
\end{equation}
as usual (e.\ g.\ \cite{Nikiforov2005}, \cite{BarShalom1989}, \cite{Ovechkin2014}) are expressed using the Slater integrals
\begin{equation}
F_{jk}^{(0)}=e^2\int_0^{\infty}\int_0^{\infty}\frac 1{x_>}R_{jj}(x_1)R_{kk}(x_2)dx_1dx_2,\quad
G_{jk}^{(s)}=e^2\int_0^{\infty}\int_0^{\infty}\frac{x_<^s}{x_>^{s+1}}R_{jk}(x_1)R_{jk}(x_2)dx_1dx_2,
\end{equation}
\begin{equation}
R_{jk}(x)=f_j(x)f_k(x)+g_j(x)g_k(x),\ x_>=\max\{x_1,x_2\},\ x_<=\min\{x_1,x_2\};
\end{equation}
the angular factor
\begin{equation}
\mathfrak{R}_{s\kap_j\kap_k}=\frac{(\kap_j+\kap_k-s)(\kap_j+\kap_k+s+1)}{\kap_j\kap_k}\cdot\wigner{s}{\ell(\kap_j)}{\ell(\kap_k)}^2,
\end{equation}
\begin{equation}
\ell(\kap)=|\kap|-\heaviside(-\kap)
\end{equation}
and
\begin{equation}
\heaviside(u)=\left[\begin{array}{ll}1,&u\geq 0\\0,&u<0\end{array}\right.,\,u\in\RR;\quad
\delta_{kl}=\left[\begin{array}{ll}1,&k=l\\0,&k\neq l\end{array}\right.,\,k,l\in\ZZ
\end{equation}
represent the Heaviside step function and the Kronecker delta, respectively. Note that the expression for $q_i$ can be modified to more accurately take into account the interaction with free electrons (see, e.~g.~\cite{Ovechkin2014}).

Consider a locally thermodynamically equilibrated plasma. Let $T$ be its temperature, $\beta = (k_{\rm B}T)^{-1}$; $\mu$ be the chemical potential, $\eta =\beta\mu$. 
The distribution over the set of configurations $\config$ is assumed to be Gibbsian, corresponding to the energies \eqref{ZeroEnergy}
\begin{equation}
P_{\bf n}=\frac{1}{\Xi(\beta,\eta)}\statves({\bf n})\exp{-\beta E^{(0)}({\bf n})+\eta\sum_{j=1}^Mn_j},
\quad \statves({\bf n})=\prod_{j=1}^M\binom{\statves_j}{n_j}
\end{equation}
\begin{equation}
\Xi(\beta,\eta)=\sum_{{\bf n}\in\config}\statves({\bf n})\exp{-\beta E^{(0)}({\bf n})+\eta\sum_{j=1}^Mn_j}
\end{equation}
what as is well-known (e.\ g.\ \cite{Carson1968}) results in the binomial distribution 
\begin{equation}\label{GibbsBinom}
\Xi(\beta,\eta)=\prod_{j=1}^M(1+\exp{-\beta\eps_j+\eta})^{\statves_j},\quad
P_{\bf n}=\prod_{j=1}^M\binom{\statves_j}{n_j}p_j^{n_j}(1-p_j)^{\statves_j-n_j}
\end{equation}
where $p_j=(1+\exp{\beta\eps_j-\eta})^{-1}$ are the occupation fractions of the shells. 

We will describe the bound-bound opacity by means of a cross-section per atomic cell. For this cross-section we start from the expression (\cite{Nikiforov2005})
\begin{equation}\label{SigmaBBInitial}
\sigma_{\rm bb}(\omega)=2\pi^2\alpha a_0\frac{e^2}{\hbar}\sum_{{\bf n}\in\config}P_{\bf n}\sum_{i,f=1}^M
n_i\left(1-n_f/\statves_f\right)\mathfrak{f}_{if}S\Bigl(\omega-E_{if}({\bf n})/\hbar;L_{if}({\bf n}),G_{if}({\bf n})\Bigr)
\end{equation}
(we exploit the usual notations $e$, $m$, $\hbar$, $c$, $a_0$, $\alpha\approx 1/137$ for electron charge, electron mass, Planck's constant, speed of light, Bohr's radius and fine structure constant, respectively).
The one-electron transition oscillator strength in \eqref{SigmaBBInitial} is calculated as (\cite{Nikiforov2005})
\begin{multline}\label{OscillatorStrength}
\mathfrak{f}_{ab}=\frac{2mc^2}{3(\eps_b-\eps_a)}\left(
\frac{\delta_{\kap_a+\kap_b,0}}{4\kap_a^2-1}+\frac{\kap_b}{\kap_a+\kap_b}\delta_{|\kap_a-\kap_b|,1}\right)\times\\\times\left(
(\kap_a-\kap_b-1)\int_0^{\infty}f_a(x)g_b(x)dx+(\kap_a-\kap_b+1)\int_0^{\infty}f_b(x)g_a(x)dx\right)^2
\end{multline}
for $a,b=1,\ldots,M$. It is important to note that the oscillator strength given by \eqref{OscillatorStrength} does not depend on configuration.

The configurations with $n_i=0$ or $n_f=\statves_f$ give no contribution to the inner sum in \eqref{SigmaBBInitial}. For the rest configurations we take the transition energy $E_{if}({\bf n})$ in \eqref{SigmaBBInitial} in the form 
\begin{equation*}
E_{if}({\bf n})=E({\bf n}-{\bf e}_i+{\bf e}_f)-E({\bf n})=
q_f-q_i+\sum_{j=1}^M(n_j-\delta_{ij})(\theta_{fj}-\theta_{ij}).
\end{equation*}
Here ${\bf n}-{\bf e}_i+{\bf e}_f$ is the configuration obtained by the transition of one electron between shells $i$ and $f$. We use the notation ${\bf e}_j$ for the vector in $\RR^M$ with coordinates $\delta_{sj}$, $s=1,\ldots,M$. 

For the lineshape function $S$ we use Voight shape. For the sake of convenience we represent it in the form of the inverse Fourier transform
\begin{equation}\label{Voight}
S(w,L,G)=\frac{1}{\pi^{3/2}G}\int_{\RR}\frac{L\exp{-\frac{u^2}{G^2}}}{(w-u)^2+L^2}du=
\frac 1{2\pi}\int_{\RR}\exp{-\ii tw-\frac{G^2t^2}4-L|t|}dt
\end{equation}
For the Gaussian component $G_{if}^2$ of \eqref{Voight} we take into account the UTA effective width (see \cite{Bauche2015}) and Doppler width
\begin{equation}
G_{if}^2({\bf n})=2\Delta^2_{if}({\bf n})+D_{if}^2,\quad D_{if}^2=\frac{2(\eps_f-\eps_i)^2}{1836Amc^2\beta\hbar^2}
\end{equation}
UTA width is taken as given in \cite{BarShalom1995}
\begin{equation}\label{UTA}
\Delta^2_{if}=\sum_{j=1}^M(n_j-\delta_{ij})(\statves_j-n_j-\delta_{fj})d_{j;if}^2.
\end{equation}
The Lorentzian component $L_{if}$ can include natural and collisional widths (e.\ g.\ \cite{Nikiforov2005}). Its dependence on the configuration ${\bf n}$ has the form 
\begin{equation}\label{Lorentz}
L_{if}({\bf n})=\sum_{1\leq a\neq b\leq M}\frac{n_a}{\statves_a}\left(1-\frac{n_b}{\statves_b}\right)\lambda_{ab;if}
\end{equation}
(see \cite{Bauche2015}, \cite{Peyrusse1999}). 
The coefficients $\lambda$, $d^2$ in \eqref{UTA}, \eqref{Lorentz} do not depend on the occupation numbers.

\section{Fourier transform representation of bound-bound\protect\\opacity}
Changing the order of summations in \eqref{SigmaBBInitial} and using the identities
\begin{equation}\label{Binom1}
n\binom{\statves}{n}p^n(1-p)^{\statves-n}=\statves p\binom{\statves-1}{n-1}p^{n-1}(1-p)^{\statves-1-(n-1)},
\end{equation}
\begin{equation}
(1-n/\statves)\binom{\statves}{n}p^n(1-p)^{\statves-n}=(1-p)\binom{\statves-1}{n}p^{n}(1-p)^{\statves-1-n}
\end{equation}
we obtain
\begin{equation}\label{SigmaBBFinal}
\sigma_{\rm bb}(\omega)=2\pi^2\alpha a_0\frac{e^2}{\hbar}\sum_{i,f=1}^M
\statves_ip_i\left(1-p_f\right)\mathfrak{f}_{if}X_{if}(\omega)
\end{equation}
where
\begin{equation}\label{XDefinition}
X_{if}(\omega)=\sum_{{\bf n}\in\config}P^{(if)}_{\bf n}S\Bigl(\omega-E_{if}({\bf n})/\hbar;L_{if}({\bf n}),G_{if}({\bf n})\Bigr)
\end{equation}
and
\begin{equation}
P^{(if)}_{\bf n}=\prod_{j=1}^M\binom{\statves_j-\delta_{j;if}}{n_j-\delta_{ji}}p_j^{n_j-\delta_{ji}}(1-p_j)^{\statves_j-\delta_{jf}-n_j}
\end{equation}
(we use the notation $\delta_{j;if}=\delta_{ji}+\delta_{jf}$). 

Let $\xi\sim{\rm Binomial}(m,p)$, $m\in\ZZ_+$, $p\in[0,1]$ mean the binomial distribution of random variable $\xi$, that is
\begin{equation}
\prob(\xi=k)=\binom{m}{k}p^k(1-p)^{m-k},k=0,\ldots,m.
\end{equation}
Introduce the random vector
\begin{equation}
\xxi_{if}=\left(\xi_{1;if},\xi_{2;if},\ldots,\xi_{M;if}\right)^{\top}
\end{equation}
with independent binomial components (e.\ g.\ \cite{Shiryaev1996})
\begin{equation}
\xi_{j;if}\sim{\rm Binomial}(\statves_j-\delta_{j;if},p_j),\ j=1,\ldots,M.
\end{equation}
Using these notations we get 
\begin{equation}\label{ProbabilisticBB}
X_{if}(\omega)=\me S\Bigl(\omega-E_{if}(\xxi_{if}+{\bf e}_i)/\hbar;L_{if}(\xxi_{if}+{\bf e}_i),G_{if}(\xxi_{if}+{\bf e}_i)\Bigr).
\end{equation}
As usual, $\prob$, $\me$ and $\var$ denote the probability of an event, the mathematical expectation and the variance of a random variable. 
Note that for $j\neq i,f$ the random variable $\xi_{j;if}$ has the same distribution as the shell occupation number $n_j$ according to \eqref{GibbsBinom}.

We will now focus on calculating $X_{if}$. To simplify the following formulas we omit indices $i,f$ for $G_{if}$, $D_{if}$, $\Delta_{if}$, $d_{j;if}$, $L_{if}$, $\lambda_{ab;if}$, $\xxi_{if}$, $\xi_{j;if}$ and denote
\begin{equation}
u=(q_f-q_i)/\hbar \mbox{ and } w_{j}=(\theta_{fj}-\theta_{ij})/\hbar.
\end{equation}
Replacing the orders of (double) integration in \eqref{ProbabilisticBB} we get
\begin{equation}\label{ProbabilisticBB:2}
X_{if}(\omega)=
\frac 1{2\pi}\int_{\RR}\me\exp{-\ii t(\omega-u-\sum_{j=1}^Mw_j\xi_j)-\frac{G^2(\xxi+{\bf e}_i)t^2}4-L(\xxi+{\bf e}_i)|t|}dt.
\end{equation}
For approximate evaluation of \eqref{ProbabilisticBB:2} we replace the Lorentz width by its average value over configurations
\begin{equation}\label{AverageLorentz}
L_0=\me L(\xxi+{\bf e}_i)=\sum_{1\leq a\neq b\leq M}\lambda_{ab}
\left(p_a+\frac{\delta_{ai}(1-p_a)-\delta_{af}p_a}{\statves_a}\right)
\left(1-p_b-\frac{\delta_{bi}(1-p_b)-\delta_{bf}p_b}{\statves_b}\right).
\end{equation}
The first two terms in the exponent in \eqref{ProbabilisticBB:2} are decomposed into the sum of statistically independent summands
\begin{equation}
-\ii t\left(\omega-u-\sum_{j=1}^Mw_j(\statves_j-\delta_{j;if})p_j\right)-\frac{D^2t^2}4+\sum_{j=1}^M\left(\ii tw_j(\xi_j-\me\xi_j)-
\frac{d_j^2t^2}2\xi_j(\statves_j-\delta_{j;if}-\xi_j)\right).
\end{equation}
Using that the average of independent random variables product is equal to the product of the averages we obtain
\begin{equation}\label{ProbabilisticBB:3}
X_{if}(\omega)=
\frac 1{2\pi}\int_{\RR}
\exp{-\ii t(\omega-\omega_{fi}^{(0)})-\frac{D^2t^2}4-L_0|t|}\cdot
\prod_{j=1}^M\Psi_j(t)dt
\end{equation}
where
\begin{equation}\label{DefOmegafi}
\omega_{fi}^{(0)}=u+\sum_{j=1}^Mw_j(\statves_j-\delta_{j;if})p_j
\end{equation}
\begin{multline}\label{DefPsij}
\Psi_j(t)=\me\exp{\ii tw_j(\xi_j-\me\xi_j)-\frac{d_j^2t^2}2\xi_j(\statves_j-\delta_{j;if}-\xi_j)}=\\
=\sum_{k=0}^{\statves_j-\delta_{j;if}}\binom{\statves_j-\delta_{j;if}}{k}p_j^k(1-p_j)^{\statves_j-\delta_{j;if}-k}
\exp{\ii tw_j(k-(\statves_j-\delta_{j;if})p_j)-\frac{d_j^2t^2}2k(\statves_j-\delta_{j;if}-k)}
\end{multline}
The calculation of $\Psi_j(t)$ on the grid of argument values sufficient to perform the Fourier transform in \eqref{ProbabilisticBB:3} appears to be significantly less expensive than direct summation of $\prod_{j=1}^M\statves_j$ terms of \eqref{XDefinition} right side. The expressions \eqref{SigmaBBFinal}, \eqref{ProbabilisticBB:3} together with \eqref{DefOmegafi}, \eqref{DefPsij} represent the main result of this work. For calculations using these expressions we will use the abbreviation DCA+UTA.

The method of approximate evaluation of $X_{if}(\omega)$ allows for some further simplifications. Namely, we can replace not only the the Lorentzian but also the squared UTA width with its average value
\begin{equation}\label{AverageUTA}
U_0^2=\me\Delta^2(\xxi+{\bf e}_i)=\sum_{j=1}^M
d_j^2(\statves_j-\delta_{j;if})(\statves_j-\delta_{j;if}-1)p_j(1-p_j).
\end{equation}
After this replacement we obtain that 
\begin{equation}\label{ProbabilisticBB:4}
X_{if}(\omega)=
\frac 1{2\pi}\int_{\RR}
\exp{-\ii t(\omega-\omega_{fi}^{(0)})-(D^2+2U_0^2)t^2/4-L_0|t|}
\cdot\Phi(t)dt
\end{equation}
where
\begin{equation}\label{DefPhi}
\Phi(t)=\prod_{j=1}^M\me\exp{\ii tw_j(\xi_j-\me\xi_j)}=
\prod_{j=1}^M\left((1-p_j)\exp{-\ii t p_jw_j}+p_j\exp{\ii t(1-p_j)w_j}\right)^{\statves_j-\delta_{j;if}}.
\end{equation}
We mark the corresponding calculations as DCA+UTAeff.

If we neglect the UTA detailed structure contribution (that corresponds to setting $U_0=0$), we obtain a simpler expression 
\begin{equation}\label{ProbabilisticBB:5}
X_{if}(\omega)=
\frac 1{2\pi}\int_{\RR}
\exp{-\ii t(\omega-\omega_{fi}^{(0)})-D^2t^2/4-L_0|t|}
\cdot\Phi(t)dt
\end{equation}
and use the abbreviation DCA for the corresponding calculations.

In \eqref{ProbabilisticBB:4} different line broadening factors are separated: Doppler, UTA, Lorentz (natural and collisional) and the last --- broadening due to the occupation numbers fluctuations. The function $\Phi(t)$ describing these fluctuations is the characteristic function (e.\ g.\ \cite{Shiryaev1996}) of centered random variable
\begin{equation}
\sum_{j=1}^Mw_j(\xi_j-\me\xi_j)
\end{equation}
with variance
\begin{equation}
\Sigma^2=\var\left(\sum_{j=1}^Mw_j(\xi_j-\me\xi_j)\right)=\sum_{j=1}^Mw_j^2(\statves_j-\delta_{j;if})p_j(1-p_j).
\end{equation}
Under certain conditions (e.\ g.\ \cite{Shiryaev1996}) the distribution of this random variable is approximately Gaussian and 
\begin{equation}
\Phi(t)\approx\exp{-\frac{\Sigma^2t^2}2}.
\end{equation}
This gives an approximation for \eqref{ProbabilisticBB} via the Voigt function
\begin{equation}\label{Effective}
X_{if}\approx S\left(\omega-\omega_{fi}^{(0)},L_0,\sqrt{D^2+2U_0^2+2\Sigma^2}\right).
\end{equation}
The last expression is similar to that obtained in \cite{Nikiforov2005}, \cite{Dragalov1987}. We mark the corresponding calculations as DCAeff+UTAeff.

\section{Calculation results and discussion}
To illustrate the described method we calculated the transmission spectra for several well-known experiments. The average atom characteristics obtained within Dirac --- Hartree --- Fock --- Slater approximation according to \cite{Nikiforov2005} were used for the calculations. To find the bounded states we applied the phase method \cite{Vronskiy2018:2}. The spectral transmission coefficient through the plasma layer is defined by usual expression
\begin{equation}
\tau(\omega)=\exp{-\sigma(\omega)L\rho N_{\rm Av}/A}
\end{equation}
where $\rho$ is plasma density, $T$ --- its temperature and $L$ --- width of the layer, $A$ stands for the atomic weight and $N_{\text{Av}}=6.022\cdot 10^{23}$ is Avogadro's number. For the total opacity cross section we use the usual expression
\begin{equation}
\sigma(\omega)=(1-\exp{-\hbar\omega/T})(\sigma_{\rm bb}(\omega)+\sigma_{\rm bf}(\omega)+\sigma_{\rm ff}(\omega))+\sigma_{\rm sc}(\omega),
\end{equation}
where the bound–bound opacity cross section $\sigma_{\rm bb}$ is determined using the expressions from this paper, and the bound–free, free–free and scattering cross sections $\sigma_{\rm bf}$, $\sigma_{\rm ff}$ and $\sigma_{\rm sc}$ are calculated in the average atom approximation according to \cite{Nikiforov2005}.

\begin{figure}
\centerline{\includegraphics[width=0.8\textwidth]{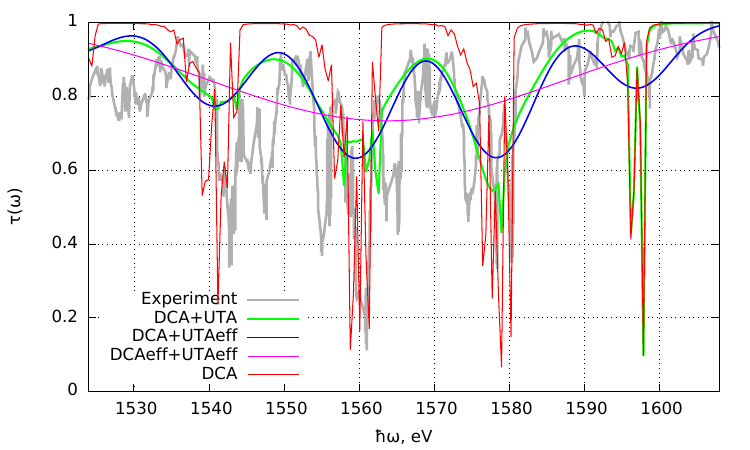}}
\caption{The calculated spectral transmission through the aluminium plasma layer at $\rho=0.02\,\text{g/cm}^3$, $T=58$ eV: DCA+UTA --- expr. \eqref{SigmaBBFinal}, \eqref{ProbabilisticBB:3}; DCA+UTAeff --- expr. \eqref{SigmaBBFinal}, \eqref{ProbabilisticBB:4}; 
DCA --- expr. \eqref{SigmaBBFinal}, \eqref{ProbabilisticBB:5}; 
DCAeff+UTAeff --- expr. \eqref{SigmaBBFinal}, \eqref{Effective} compared to the experiment \cite{Perry2000})}\label{figAl}
\end{figure}

In Fig.~\ref{figAl} we compare of the calculated $\tau(\omega)$ for the aluminium plasma with density $\rho=0.02\,\gccm$, temperature $T=58$ eV and $L\rho=1.35\cdot 10^{-5}\,\gscm$ with the experimental transmission taken from \cite{Perry2000}. 
Calculation of $\tau(\omega)$ using our main formulas \eqref{SigmaBBFinal}, \eqref{ProbabilisticBB:3} agrees satisfactorily with the experimental data on the positions of transition groups and their widths (except for the group of transitions with energies near 1.6 keV). The shift in energies compared to the experimental one is due to the use of characteristics of the average atom. The simplified formulas \eqref{SigmaBBFinal}, \eqref{ProbabilisticBB:4} with UTA widths averaged over the configurations give results that agree with more accurate calculation, except for the group of transitions with energies near 1.6 keV, formed mainly due to transitions from He-like ions (for which the UTA width is small or equal to zero, and differs essentially from the average over the configurations). The Fourier transform method of DCA with neglecting the UTA widths according to \eqref{SigmaBBFinal}, \eqref{ProbabilisticBB:5} illustrates the effect of the shells occupation numbers fluctuations on the line groups shapes. Ignoring the detailed structure of the transition leads to a significant underestimation of the width of the transition groups compared to the experimental ones. Effective accounting of the occupation numbers fluctuations according to the formulas \eqref{SigmaBBFinal}, \eqref{Effective} poorly describes the experimental spectra.

\begin{figure}
\centerline{\includegraphics[width=0.8\textwidth]{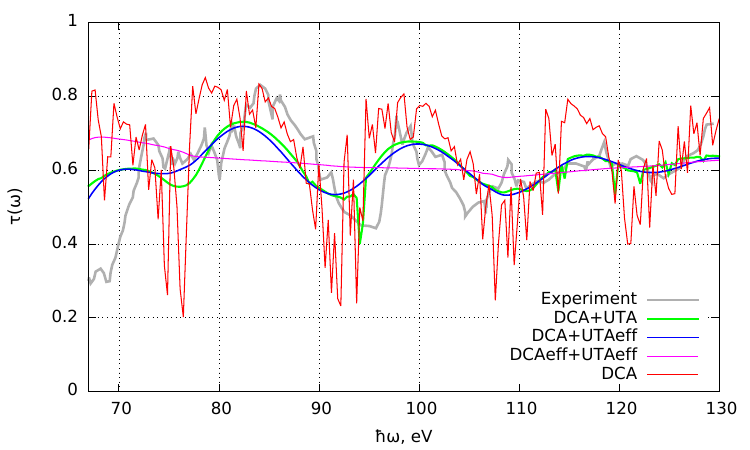}}
\caption{The spectral transmission through the iron plasma layer at $\rho=0.01\,\text{g/cm}^3$, $T=22$ eV: DCA+UTA --- expr. \eqref{SigmaBBFinal}, \eqref{ProbabilisticBB:3}; DCA+UTAeff --- expr. \eqref{SigmaBBFinal}, \eqref{ProbabilisticBB:4}; 
DCA --- expr. \eqref{SigmaBBFinal}, \eqref{ProbabilisticBB:5}; 
DCAeff+UTAeff --- expr. \eqref{SigmaBBFinal}, \eqref{Effective} compared to the experiment \cite{Winhart1996}}\label{figFe}
\end{figure}

\begin{figure}
\centerline{\includegraphics[width=0.8\textwidth]{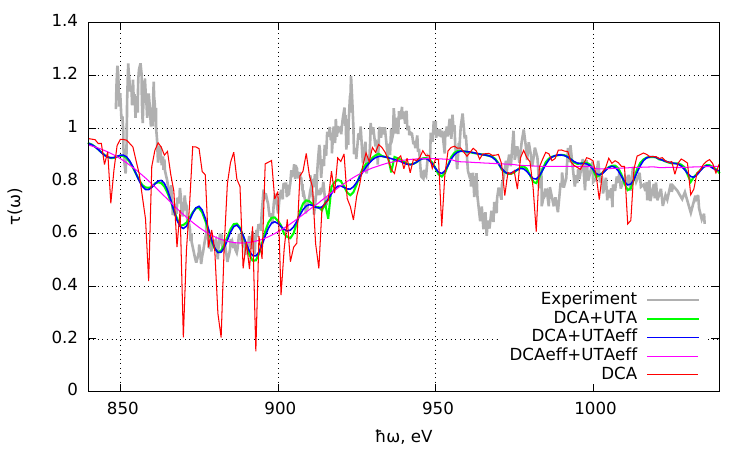}}
\caption{The spectral transmission through the nickel plasma layer at $\rho=0.01\,\text{g/cm}^3$, $T=19$ eV: DCA+UTA --- expr. \eqref{SigmaBBFinal}, \eqref{ProbabilisticBB:3}; DCA+UTAeff --- expr. \eqref{SigmaBBFinal}, \eqref{ProbabilisticBB:4}; 
DCA --- expr. \eqref{SigmaBBFinal}, \eqref{ProbabilisticBB:5}; 
DCAeff+UTAeff --- expr. \eqref{SigmaBBFinal}, \eqref{Effective} compared to the experiment \cite{Pain2021}}\label{figNi}
\end{figure}

\begin{figure}
\centerline{\includegraphics[width=0.8\textwidth]{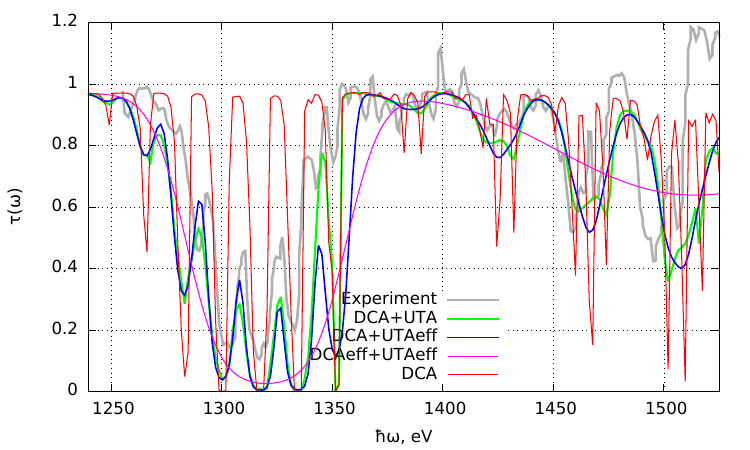}}
\caption{The spectral transmission through the magnesium plasma layer at $\rho=0.012\,\text{g/cm}^3$, $T=45$ eV: DCA+UTA --- expr. \eqref{SigmaBBFinal}, \eqref{ProbabilisticBB:3}; DCA+UTAeff --- expr. \eqref{SigmaBBFinal}, \eqref{ProbabilisticBB:4}; 
DCA --- expr. \eqref{SigmaBBFinal}, \eqref{ProbabilisticBB:5}; 
DCAeff+UTAeff --- expr. \eqref{SigmaBBFinal}, \eqref{Effective} compared to the experiment \cite{Renaudin2006}}\label{figMg}
\end{figure}

Fig.~\ref{figFe} demonstrates the similar dependencies for the experiment with iron transmission (with density $\rho=0.01\,\gccm$, temperature $T=22$ eV and $L\rho=1.5\cdot 10^{-5}\,\gscm$) from \cite{Winhart1996}. 
In Fig.~\ref{figNi} we present calculated spectra of transmission through the nickel plasma layer (with density $\rho=0.01\,\gccm$,  temperature $T=19$ eV and $L\rho=2\cdot 10^{-5}\,\gscm$) and the corresponding experimental data from \cite{Pain2021}. Finally, Fig.~\ref{figMg} shows the comparison of the calculated $\tau(\omega)$ for the magnesium plasma at density $\rho=0.012\,\gccm$, temperature $T=45$ eV and $L\rho=1.356\cdot 10^{-4}\,\gscm$ with the experimental spectra from \cite{Renaudin2006}. 

For the conditions used to obtain the data for Fig. ~\ref{figNi}, \ref{figMg}, the effective account of shell occupations fluctuations \eqref{SigmaBBFinal}, \eqref{Effective} satisfactorily describes the experimental groups of lines in the soft part of the spectrum. 
Also, for these examples the calculated spread of lines is overestimated comparing to experimental one. The reason of this overestimation is the known fact that the binomial distribution over configurations \eqref{GibbsBinom} increases the dispersion of the ion charge compared to the Gibbs distribution with first-order energy of configuration \eqref{FirstEnergy}.

\section*{Conclusions}

In this work the Fourier transform method of the Detailed Configuration Accounting for the bound-bound opacity calculations is presented. It extends the Hazak --- Kurzweil method of STA opacity spectra configurational resolution. The final formulas for the bound-bound opacity cross sections are represented as Fourier transforms of explicitly given expressions. Several simplifying approximations are also presented.
To illustrate the method we used an average atom model with Dirac one-electron states. At the same time, the method is also applicable for another ways to describe one-electron states: Schrödinger orbitals, parametric models, etc. The key assumption for the presented method is the factorization of the probabilities of configurations (i.e., the statistical independence of the shell occupation numbers). The considered approximation of binomial distribution is known to somewhat widthen the transition line groups compared to the case with the interaction taken into account. 
The proposed expressions may include configurations with a total number of electrons exceeding the charge of the nucleus, but the contribution of such configurations is insignificant.

\section*{Acknowledgments}
The authors are grateful to A.~A.~Ovechkin and E.~S.~Tsoy for helpful discussions.

\end{document}